\documentclass{ieeeaccess}
\usepackage{cite}
\usepackage{amsmath,amssymb,amsfonts}
\usepackage{algorithmic}
\usepackage{graphicx}
\usepackage{textcomp}
\def\BibTeX{{\rm B\kern-.05em{\sc i\kern-.025em b}\kern-.08em
    T\kern-.1667em\lower.7ex\hbox{E}\kern-.125emX}}
\usepackage{bm}
\usepackage{mathtools}
\usepackage{etoolbox}
\usepackage{soul}
\usepackage{xcolor}
\usepackage{siunitx}
\usepackage[switch, modulo]{lineno}
\usepackage[colorlinks=true, linkcolor=blue, citecolor=blue, urlcolor=blue]{hyperref}

\soulregister{\textsubscript}{1}
\soulregister{\textsuperscript}{1}
\soulregister{\cite}{7}
\soulregister{\label}{7}
\soulregister{\caption}{1}
\soulregister{\ref}{7}
\soulregister{\eqref}{7}
\soulregister{\textit}{1}
\soulregister{\emph}{1}
\soulregister{\mu}{0}
\newcommand{\hlcite}[1]{\mbox{\cite{#1}}}
\newcommand{\hlref}[1]{\mbox{\ref{#1}}}
\soulregister{\hlcite}{7}
\soulregister{\hlref}{7}
\newcommand{\RevTwo}[2]{#1}
\newcommand{\RevThree}[2]{#1}
\newcommand{\RevTwoCap}[2]{#1}
\newcommand{\ReviewSpan}[2]{#2}

\makeatletter
\def\ps@titlepage{%
  \def\@oddhead{}%
  \def\@evenhead{}%
  \def\@oddfoot{\hbox{}\hfil\thepage}%
  \def\@evenfoot{\thepage\hfil\hbox{}}%
}

\def\ps@headings{%
  \def\@oddhead{\hbox{}\hfil\thepage}%
  \def\@evenhead{\thepage\hfil\hbox{}}%
  \def\@oddfoot{}%
  \def\@evenfoot{}%
}
\makeatother
\begin{document}
\history{This document has not been peer reviewed but has been cleared by NIST for release. Date of publication xxxx 00, 0000, date of current version 11/19/2026. }
\doi{10.1109/TQE.2020.DOI}

\title{Physical Learning for Autonomous Tuning of Spin Qubits in Electrostatically Defined Quantum Dots}
\author{\uppercase{Leonardo Gaytan}\authorrefmark{1,$\dagger$},
\uppercase{Bradley Lloyd}\authorrefmark{1,$\dagger$},
\uppercase{Paul Kliewer}\authorrefmark{1},
\uppercase{Chandler Wilburn}\authorrefmark{1},
\uppercase{Lincoln D. Carr}\authorrefmark{1,2,3},
\uppercase{Adam McCaughan}\authorrefmark{4},
\uppercase{Sonia Buckley}\authorrefmark{4}, and
\uppercase{Meenakshi Singh}\authorrefmark{1,2}}
\address[1]{Department of Physics, Colorado School of Mines, Golden, CO 80401 USA}
\address[2]{Quantum Engineering Program, Colorado School of Mines, Golden, CO 80401 USA}
\address[3]{Institute of Science and Technology Austria (ISTA), 3400 Klosterneuburg, Austria}
\address[4]{National Institute of Standards and Technology, Boulder, CO 80305 USA}
\address[$\dagger$]{These authors contributed equally to this work.}
\tfootnote{This work was supported by the National Science Foundation under Award DMR-2046428.
The work of B. Lloyd was supported by an NRT fellowship under Award DGE-2125899. This work was initiated, in part, at the Aspen Center for Physics, which is supported by National Science Foundation grant PHY-2210452.}

\markboth
{Gaytan \headeretal: Physical Learning for Autonomous Tuning of Spin Qubits}
{Gaytan \headeretal: Physical Learning for Autonomous Tuning of Spin Qubits}

\corresp{Corresponding author:Meenakshi Singh (email: msingh@mines.edu).}

\begin{abstract}
Tuning electrostatically defined quantum dots remains a primary bottleneck in the scalability of semiconductor spin qubits. While existing solutions rely on machine learning algorithms, these approaches require global information, introduce latency bottlenecks, and tax the limited cooling power available at cryogenic stages. Here, we demonstrate an FPGA-in-the-loop learning paradigm that bypasses digital computing overhead by utilizing multiplexed gradient descent (MGD) for tuning. Implemented on an FPGA and an emulated quantum dot device, our system autonomously tracks environmental drift, dynamically recovers from discrete charge jumps, and maintains steady-state stabilization in low signal-to-noise ratio regimes, without prior knowledge of the device's cross-capacitance matrix. The system also navigates the charge-stability landscape to stabilize the quantum dot around optimal operating points. These results demonstrate a ``compute-in-place'' platform, with a logic and power footprint suited to future cryo-CMOS integration, that provides autonomous stabilization of quantum dots at desired operating points, creating an auto-tuning system for spin qubits that is resilient to electrostatic drifts, changing noise profiles, and two-level fluctuations.
\end{abstract}

\begin{keywords}
Charge noise, field programmable gate arrays, gradient methods, quantum dots,
semiconductor device modeling, spin qubits, stochastic approximation
\end{keywords}

\titlepgskip=-15pt

\maketitle

\section{Introduction}
\label{sec:intro}

\PARstart{S}{pin} qubits in electrostatically defined quantum dots are among the most promising candidates for a scalable solid-state quantum computing platform, owing in large part to their compatibility with existing semiconductor fabrication infrastructure and their potential for dense on-chip integration \cite{loss_quantum_1998}. While coherence times in isotopically purified silicon platforms have advanced substantially, with recent demonstrations of gate fidelities exceeding 99.9\%, charge noise remains a persistent and fundamental obstacle to high-fidelity operation \cite{yoneda_quantum-dot_2018}. Thus, tuning the spin qubits to optimal operating points and stabilizing them at those points over long periods of time remain high-reward challenges. Moreover, the challenge extends well beyond noise mitigation for individual devices: scaling to the thousands or millions of physical qubits required for fault-tolerant computation demands that every qubit be autonomously initialized, tuned, and stabilized in real time.

Existing automation spans a hierarchy of tasks \cite{zwolak_colloquium_2023, taylor_spin_2026}: coarse tuning into the few-electron regime \cite{moon_machine_2020,baart_computer-automated_2016,botzem_tuning_2018}, charge state identification \cite{kalantre_machine_2019,durrer_automated_2020}, cross-capacitance compensation via virtual gates \cite{mills_computer-automated_2019,volk_loading_2019}, and fine tuning of qubit parameters \cite{teske_machine_2019, van_diepen_automated_2018}, culminating in end-to-end autonomous tuning of a spin qubit \cite{schuff_fully_2026}. These methods are designed to reach an operating point once, typically from two-dimensional scans of the charge stability diagram whose acquisition time grows with the number of gate pairs. Once reached, however, the operating point is not stationary: 1/f charge noise and TLS switching displace it continuously \cite{rudolph_long-term_2019,rao_towards_2025}. Repeating global tuning for every displacement is impractical, which motivates continuous, local stabilization. 

Continuous stabilization has recently been demonstrated in a ten-dot array by repeatedly acquiring charge stability diagrams, locating the target charge cell with trained neural networks and a Hough transform, and applying proportional corrections \cite{rao_towards_2025}. Each update, however, requires a set of 2D scans whose acquisition takes seconds and grows with the number of dot pairs, followed by image-based inference. At the other extreme, a single charge sensor has been held on a Coulomb-peak flank by a microsecond-scale PID loop \cite{nakajima_real-time_2021}. Such loops act directly on the sensor signal, which provides a signed error only on a monotonic flank. At an extremum such as an idling point, the error is symmetric about the target and carries no directional information. Independent loops also require choosing which sensor drives which gate, and with what sign. With strong cross talk, the correct choice depends on the cross-capacitance matrix, and the wrong one makes the loops diverge.


Meanwhile, physical learning methods have recently shown promise as an alternative for optimizing complex systems without the overhead of traditional processors \cite{dillavou_machine_2024, krstic_stability_2000,liu_stochastic_2012,scheinker_100_2024}. Rather than focusing on training neural networks to recognize patterns, these methods utilize hardware-native optimization methods for adaptive control which are already able to efficiently scale to large system sizes while requiring minimal computation \cite{mccaughan_multiplexed_2023,oripov_scaling_2025}. Such an approach is well-suited for electrostatically defined quantum dots because it is model-free, requires no prior training, is inherently device-agnostic, and can be implemented \textit{in situ}. By leveraging physical learning for adaptive control, this framework bypasses the data-heavy training bottlenecks of traditional machine learning while also not being constrained by predefined feedback loops between individual sensors and gates.

In this work, we demonstrate a method to locally stabilize a quantum dot system at specific operating points in the charge stability diagram using a multiplexed form of simultaneous perturbation stochastic approximation (SPSA) \cite{spall_stochastic_1987}, an algorithm that can be implemented physically in the multiplexed gradient descent (MGD) \ReviewSpan{S1}{framework} \cite{mccaughan_multiplexed_2023}. We hereafter refer to this implementation simply as MGD. MGD is a hardware-efficient, model-free optimization algorithm, which applies discrete, stochastic voltage perturbations to autonomously navigate the parameter space. Because the system relies solely on local gradient information rather than global stability maps, it can quickly probe changes in the electrostatic environment and apply autonomous corrections. To validate this framework, the target quantum dot and its inherent noise profiles, including $1/f$ drift and two-level fluctuations, were mathematically simulated, while the physical learning circuit was deployed on a room-temperature FPGA.

\section{Autonomous Tuning Architecture and Emulation Framework}
\label{sec:methods}

\subsection{Quantum Dots and the Charge Stability Diagram}


To physically realize a spin qubit, individual electrons must be trapped and isolated. In electrostatically defined quantum dots, this trapping is achieved by applying specific DC voltages to an array of nanoscale metallic gates \RevTwo{(Fig.~\ref{fig:charge_stability_diagram}a)}{S2}. These gate voltages form a continuous parameter space that is mapped to a discrete set of charge configurations via the charge stability diagram. For operation as a spin qubit, the system must be tuned to a specific charge configuration within this multidimensional voltage space. \RevTwo{As shown in Fig.~\ref{fig:charge_stability_diagram}b}{S3}, the charge stability diagram consists of charge-occupancy `honeycomb cells' each corresponding to a specific electron number. 


While the charge configuration within a cell is consistent, specific points in the cell will be more desirable than others. In a singlet-triplet (S-T) qubit configuration, for example, a position deep in (1,1) where the gradient of the coupling $J(\epsilon)$ is small may be chosen while the qubit is ``idling'' to isolate from charge noise. Conversely, when it is time to pulse the qubit, the position in the charge stability diagram might be moved closer to the inter-dot transition where the gradient of $J(\epsilon)$ is large, making it easier to pulse the qubit at the expense of greater coupling to charge noise. These optimal control points will hereafter be referred to as `parking points'.

Regardless of the chosen point, the charge stability diagram is not stationary. Physical devices suffer from $1/f$ pink noise, which causes the entire charge stability diagram to continuously drift, as well as Two-Level Systems (TLS) that cause sudden, discrete spatial jumps in the optimal coordinate. Together, these effects mean that any viable control system must track a continuously moving and occasionally jumping target in order to remain at the desired operating point. 

Here we propose a physical learning method that can (i) automatically find and maintain position around an idling point in the cell and (ii) automatically maintain position at an ideal pulsing parking point, previously identified by conventional means not discussed here.

\Figure[htbp](topskip=0pt, botskip=0pt, midskip=0pt)[width=6.1in]{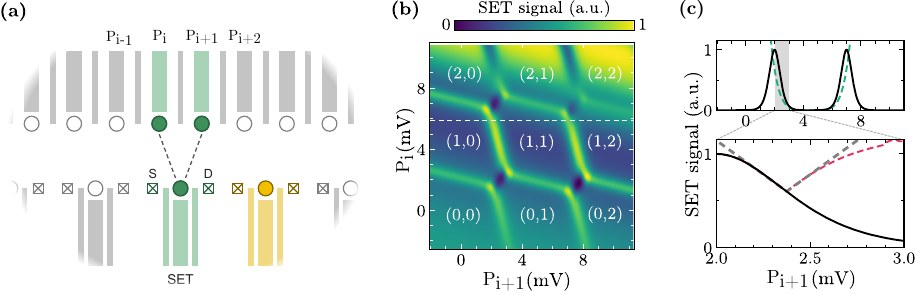}
{\textbf{Charge stability diagram and cost-function landscape.}
    (a) Gate layout, showing the two dots and SET (green) and grayed out gates that are not associated with measured quantum dots of their own but still have crosstalk with the measured dots. \ReviewSpan{S5}{A second SET (yellow) is required in order to balance the configuration on the flank of a charge transition peak.}
    (b) Simulated \cite{hader_simulation_2024} charge stability diagram (dI/dV) of a double quantum dot, showing the
    normalized SET signal as a function of the left and right gate voltages,
    \RevTwoCap{$P_i$ and $P_{i+1}$}{S6}. Dark regions are cells of stable charge occupation, labeled
    $(N_L, N_R)$ by the number of electrons in the left and right dots, and the
    bright lines are the charge transitions that separate them. The white dashed
    line marks the cut plotted in \RevTwoCap{(c)}{S10}.
    \ReviewSpan{S7}{(c)} Top: \ReviewSpan{S8}{SET signal} as a function of \RevTwoCap{$P_{i+1}$}{S9} along the dashed line in \RevTwoCap{(b)}{S10},
    showing two transition peaks separated by a region of fixed charge occupation.
    The green dashed curve is the polynomial cost function of
    \eqref{eq:idle}, used to navigate to and hold around the idling point at the
    center of the cell. Bottom: enlarged view of the flank of a transition peak. The
    gray dashed curve is the cost function of \eqref{eq:pulse}, used to hold a pulsing point on the steep edge of the peak. \RevTwoCap{The red dashed line represents the exact result of taking the absolute value about the pulsing point.}{S11}\label{fig:charge_stability_diagram}}
\subsection{Emulation Framework}

In order to utilize MGD methods to tune the position within the cell, we require a cost function whose gradient the algorithm can follow in order to find optimal parking points. Here, we introduce a \RevTwo{computer-}{S4}emulation of the sensor response of a single proximal single electron transistor (SET) whose signal is demodulated via a lock-in amplifier resulting in conductance ($dI_\text{SET}/dV$) peaks. In the vicinity of the double dot are other gates besides the plunger gates of the two dots. These gates could be plunger gates of other dots in the array, barrier gates, screening gates, or flanking gates. Regardless of the number or nature of these gates, the SET is assumed to only be affected by the chemical potentials of the two quantum dots of the double dots themselves, and these dots are taken to be in the middle of the array they reside in. 

We assume that the quantum dot system has already been tuned into the desired charge occupancy by other means. The emulation faithfully mirrors the analog signal path of a real device, and we utilize the \ReviewSpan{S12}{SET conductance}, normalized to peak strength \RevTwo{and unitless}{S11}, as our cost function (Fig.~\ref{fig:charge_stability_diagram}\RevTwo{c}{S7}):

\begin{equation}
    C(\mathbf{V_k},t_k) = \tilde{C}(\mathbf{V}_k,t_k)+\xi(t_k)
\end{equation}

\noindent where $t_k=k\Delta t$ is the time of the k\textsuperscript{th} iteration and $\mathbf{V}_k$ is a $N_\text{gates}$ dimensional column vector representing the physical gate voltage configuration of the k\textsuperscript{th} iteration. $\tilde{C}(\mathbf{V}_k,t_k)$ represents the cost associated with the ideal, noiseless signal and can take different forms depending on the parking point for which the algorithm is being used. $\xi(t_k)$ represents the time-varying disturbances that are superimposed on the background cost function in order to faithfully reproduce the noise characteristics of a physical device.

In the case of an idling parking point, $\tilde{C}(\mathbf{V}_k,t_k)$ will take the form of a series expansion of thermally broadened transition peaks at a temperature T=0.3K \cite{van_houten_coulomb-blockade_1992}, whereas for a pulsing point it will take the form of a simple change in signal (Fig.~\ref{fig:charge_stability_diagram}\RevTwo{c}{S7}): 

\begin{align}
    \tilde{C}_\text{idle}(\mathbf{V}_k, t_k) = \sum_{j=1}^{N_\text{poly}} \beta_j ||\mathbf{A}\left(\mathbf{V}_k-\mathbf{V}_\text{opt}(t_k)\right)||_{2j}^{2j} \label{eq:idle}\\
    \tilde{C}_\text{pulse}(\mathbf{V}_k,t_k) = \beta_\text{p}||\mathbf{A}\left(\mathbf{V}_k-\mathbf{V}_\text{opt}(t_k)\right)||\label{eq:pulse}
\end{align}

\noindent where the $\beta_j$ coefficients weight the importance of the polynomials in the idling case, $\beta_p$ describes the local slope of the transition peak at the pulsing point, $\mathbf{A}$ is a $2\times N_\text{gates}$ cross-capacitance matrix, $\mathbf{V}_\text{opt}(t_k)$ describes the position of the relevant parking point in voltage space, which is subject to time dependent drifts and jumps, and $2N_\text{poly}$ represents the order of the polynomial. \ReviewSpan{S13}{With a single SET, the pulsing cost function can only be constrained in a single direction making it a trough. Future work will use two SETs to constrain in both virtual gate directions, for now we proceed under the assumption the both directions can be constrained.}

The inclusion of the cross capacitance matrix $\mathbf{A}$ in the polynomial approximation accounts for the cross-talk between gate voltages and the chemical potentials in the double dot. $\mathbf{A}$ enters only through the emulation; it is never provided to the controller. The matrix elements of $\mathbf{A}$ take the form $A_{ij}=e^{-|i-j|/\gamma}$ \cite{volk_loading_2019}, where $\gamma$ sets the falloff of the cross-talk. Because the rank of $A$ is constrained to 2 and its elements are subject to an exponential falloff, there is expected to be little to no increase in runtime with increasing $N_\text{gates}$. Rather, increasing $\gamma$ increases the degree to which the problem is ill-conditioned, described by the condition number $\kappa$ of the Hessian $\mathbf{H}$:
\begin{align}
    \mathbf{H}(\mathbf{V}_\text{opt}(t_k)) = 2\beta_1 \mathbf{A}^\top \mathbf{A}\\
    \kappa = \frac{|\lambda_\text{max}|}{|\lambda_\text{min}|}
\end{align}
\noindent where $\kappa$ is the magnitude ratio of the largest to the smallest non-zero eigenvalue of $\mathbf{A}^\top\mathbf{A}$ and serves as our measure of ill-conditioning. Because $\mathbf{A}^\top \mathbf{A}$ will only ever have two non-zero eigenvalues, \RevTwo{convergence is evaluated as the distance to the target in}{S14} the two-dimensional chemical potential space -- described by the two \ReviewSpan{S15}{meaningful virtual gates $\tilde{\mathbf{V}}_k = \mathbf{A}\mathbf{V}_k$} -- despite the algorithm acting directly on the full $N_\text{gates}$ dimensional physical gate space.

The term $\xi(t_k)$, the time-varying disturbances superimposed on the background cost function, is given by:





\begin{equation}
    \xi(t_k)=w_k+\sum_{n=0}^{N_\text{oct}-1}u^{(n)}_{\left\lfloor k/2^n\right\rfloor}
\end{equation}
\noindent where $w_k\sim\mathcal{N}\!\left(0,\sigma_w^2\right)$ is a white-noise component and
$u^{(n)}_m\sim\mathcal{U}(-a,a)$ are the draws of a 1/f component with $N_\text{oct}$ octaves.
The white noise models instantaneous measurement noise. The 1/f contribution follows the
Voss-McCartney construction: for each octave $n$, the uniform random draws are independent and
identically distributed (i.i.d.), and the octaves are mutually independent. Octave $n$ refreshes
its value only when the quotient $\left\lfloor k/2^n\right\rfloor$ increments, thus holding its
value for $2^n$ consecutive samples so that each octave updates at half the rate of the one below
it. With equal amplitude $a$ across octaves, the sum approximates the power spectrum $S_0/f$ over
the band fixed by the sample rate and $N_\text{oct}$ \cite{voss_1f_1978}. The noise parameters are
\begin{align}
    \sigma_w^2 &= \frac{\beta^2_{\text{max}}}{\alpha^2_\text{lever}}\frac{S_0}{f_0}\Delta f,\\
    a &= \frac{\beta_{\text{max}}}{\alpha_\text{lever}}
         \sqrt{\frac{3\ln\left(2^{N_\text{oct}}\Delta t\Delta f\right)S_0}{N_\text{oct}}},
\end{align}
\noindent where $\Delta f$ is the equivalent noise bandwidth of a hypothetical lock-in amplifier,
$\alpha_\text{lever}$ is the quantum dot's lever arm, and $\beta_\text{max}$ describes the maximal
slope of the cost function with respect to voltage, where charge noise is typically quantified.
Together, these contributions reproduce the effect of continuous background charge fluctuations
on the sensor signal\cite{paladino_mathbsf1mathbsfitf_2014, dial_charge_2013}.

We characterize the overall noise
by the amplitude signal-to-noise ratio, the ratio of the peak signal to the total noise RMS,
\begin{equation}
    \mathrm{SNR} = \frac{1}{\sqrt{\sigma_w^2 + N_\text{oct}\,a^2/3}}.
\end{equation}
The parameters chosen for this work are given in Table~\ref{tab:paramTable}. They give
$\sigma_w = 1.8\times10^{-3}$ and $a = 2.3\times10^{-2}$, a total noise RMS of 6\% of the
transition peak measured by the SET, or $\mathrm{SNR} = 16.7$, generated by charge noise commensurate with real
devices \cite{connors_charge-noise_2022, paquelet_wuetz_reducing_2023, elsayed_low_2024,
rudolph_long-term_2019}. The resulting noisy scalar cost signal is what the FPGA running the
control algorithm receives at each iteration, and it is the only information the algorithm has
access to.

\begin{table}[b]
\caption{Cost Function Parameters}
\label{tab:paramTable}
\centering
\setlength{\tabcolsep}{8pt}
\begin{tabular}{lcc}
\hline
Parameter & Value & Unit\\
\hline
$\beta_p$ & 1 & mV$^{-1}$\\
$\beta_1$ & 0.0029$^{\mathrm{a}}$ & mV$^{-2}$\\
$\beta_2$ & 0.0039$^{\mathrm{a}}$ & mV$^{-4}$\\
$\beta_3$ & 0.0021$^{\mathrm{a}}$ & mV$^{-6}$\\
$N_\text{poly}$ & 3 & --\\
$N_\text{oct}$ & 20 & --\\
$f_0$ & 40 & kHz\\
$\Delta f$ & 500 & Hz\\
$\Delta t$ & 1 & ms\\
$\alpha_\text{lever}$ & 0.1$^{\mathrm{b}}$ & eV/V\\
$\beta_\text{max}$ & 1.5 & mV$^{-1}$\\
$\sqrt{S_0}$$^{\mathrm{c}}$ & 1.10 & $\mu$eV\\
$\epsilon$ & 0.5 & mV\\
\hline
\multicolumn{3}{l}{$^{\mathrm{a}}$$T = 0.3$\,K \cite{van_houten_coulomb-blockade_1992}.
  $^{\mathrm{b}}$From \cite{koch_industrial_2025,paquelet_wuetz_atomic_2022}.
  $^{\mathrm{c}}$Single-sided.}
\end{tabular}
\end{table}

\subsection{The Multiplexed Gradient Descent Algorithm}

\RevTwo{To autonomously navigate the gate voltage vector to the minimum of the noisy, computer-generated scalar cost of the previous section, we introduce an FPGA that takes the cost function as its only input and outputs gate voltages.}{S4} To keep the system at an optimal operating point in the charge stability diagram, we utilize \RevTwo{multiplexed gradient descent}{S16} (MGD). MGD bypasses the need for offline dataset processing by operating entirely in situ, utilizing only the locally provided cost function. By restricting its focus to probing the gradient of the cost function in its immediate neighborhood, MGD bypasses the latency and memory bottlenecks of traditional machine learning algorithms that must process the entire charge stability diagram.

At its core, MGD is an autonomous hardware loop that, at each iteration, probes only the local gradient in the immediate neighborhood of its current position, making it inherently agnostic to device-to-device variability and robust to the landscape shifts induced by drift and TLS events. For the quantum dot application, from a baseline voltage state, the system applies a tiny, simultaneous, and randomized voltage perturbation to every gate. By evaluating the system's cost function at two mirror-image perturbed points around the baseline and comparing them, the algorithm approximates the local gradient. The standard gradient descent update rule, defined below, is then applied

\begin{equation}
    \mathbf{V}_{k+1} = \mathbf{V}_k - \eta\,\hat{\mathbf{g}}_k, \qquad \hat{\mathbf{g}}_k = \boldsymbol{\delta}_k\,\frac{\Delta C_k}{2\epsilon},
\end{equation}

where $\mathbf{V}_k$ is the gate voltage vector at iteration $k$, $\hat{\mathbf{g}}_k$ is the approximated local gradient, $\eta$ is the step size, and $\varepsilon$ is the perturbation size.

Since MGD is a generalization of perturbative algorithms such as finite-difference and simultaneous-perturbation stochastic approximation (SPSA), it can be optimized for the specific hardware used here. In our configuration, rather than computing the exact gradient using power-hungry floating-point multiplication, MGD computes the update step using native binary bit-shifting. The gradient term is approximated by a central difference of two cost measurements taken at mirror-image points around the baseline,
\begin{equation}
    \Delta C_k = C(\mathbf{V}_k + \epsilon\boldsymbol{\delta}_k) - C(\mathbf{V}_k - \epsilon\boldsymbol{\delta}_k),
\end{equation}

where $\boldsymbol{\delta}_k \in \{-1,+1\}^{N}$ holds the random perturbation sign of each gate. \RevThree{Each sign is drawn by a 50/50 pseudo-random coin flip implemented on the FPGA with a linear-feedback shift register (LFSR); the random signs sample the high-dimensional space without correlation, avoid periodic limit cycles, and make the time-averaged update approximate the true local gradient~\hlcite{spall_stochastic_1987}. The two costs are measured in sequence, first $C_- = C(\mathbf{V}_k - \epsilon\boldsymbol{\delta}_k)$ and then $C_+ = C(\mathbf{V}_k + \epsilon\boldsymbol{\delta}_k)$, each with its own noise sample; the baseline $\mathbf{V}_k$ itself is never measured, as it is the midpoint of the two perturbed points.}{S18} Unlike a one-sided difference, the central difference cancels the curvature of the cost to leading order, so near the optimum the update responds only to the local slope, which makes it more robust to measurement noise. By constraining the learning rate $\frac{\eta}{2\epsilon}$ to inverse powers of 2, such that $\frac{\eta}{2\epsilon} = 2^{-n}$, the hardware update equation becomes

\begin{equation}
    \mathbf{V}_{k+1} = \mathbf{V}_{k} - \boldsymbol{\delta}_k\,\frac{\Delta C_k}{2^n},
\end{equation}

evaluated in the FPGA's fixed-point units.

In digital logic, division by $2^n$ is executed natively as a binary right bit-shift, and $n$ is the parameter we change to set the hardware learning rate. The update therefore contains no multiplication: the per-gate step is the cost difference, shifted right by $n$ and signed by the perturbation direction. No Digital Signal Processing (DSP) slices or floating-point arithmetic are used, which allows the system to step down the local slope of the cost landscape at less hardware cost.  The cost is transmitted to the FPGA in Q6.26 fixed point (6 integer and 26 fractional bits, normalized so the peak signal is 1), and the gate voltages are held in Q12.20 (in mV), so the shifted cost difference is added directly to the voltage register. Because the update is computed in fixed point, a cost difference too small to survive the shift produces no step. For $n = 11$ this truncation threshold corresponds to $3\times10^{-5}$ of the peak signal, about three orders of magnitude below the RMS noise of the measured difference at the Table~\hlref{tab:paramTable} noise level ($3\times10^{-2}$ of the peak), so the truncation never acts as a deadband and does not limit the precision of the controller.

Calculating the finite-difference gradient requires only two measurements per iteration, one at each of the two perturbed points $\mathbf{V}_k \pm \epsilon\boldsymbol{\delta}_k$. Consequently, $N$ iterations correspond to only $2N$ discrete voltage operations per gate.

\RevThree{In this work the quantum-dot emulation of Section~II-B runs in Python on a host computer, while the MGD controller runs entirely on the FPGA. The two are connected by a serial link: for each measurement the host sends the FPGA a single 32-bit cost value, and the FPGA returns the gate voltages to apply next; the FPGA receives no other information about the device. On a physical device, the emulated cost would be replaced by the digitized SET signal and the returned voltages would drive the gate DACs, so the same FPGA controller could be connected to a real device with minimal changes.}{S17}

\ReviewSpan{S17}{The 10-gate controller occupies 1,539} logic elements (7\% of a Cyclone IV EP4CE22), 998 registers, and no block memory or hardware multipliers; the MGD core itself uses 1,005 logic elements and 720 registers. The 100-gate build used for the scaling measurements of Section~\hlref{sec:scaling} occupies 12,973 logic elements (58\% of the same device) and 6,785 registers, again with no block memory or multipliers, so the controller grows by about 127 logic elements and 64 registers, two 32-bit voltage registers, per gate. The 10-gate build closes timing at 98.0~MHz against the 50~MHz system clock, and the logic adds only a few clock cycles (tens of nanoseconds) per iteration, negligible next to the measurement time. The Quartus power analyzer estimates 10.0~mW of core dynamic power at 50~MHz, of which the MGD core accounts for 4.8~mW; the remaining 77.6~mW of static and 27.8~mW of I/O power come from the general-purpose device and development board rather than from the algorithm. Cryogenic control electronics are limited to a power budget of a few watts at the 4~K stage~\cite{Homulle2017}, so the logic needed for MGD-based tuning uses well under 1\% of that budget. We present this as an estimate of suitability; operating the controller at cryogenic temperature has not been demonstrated here, although commercial FPGAs have been shown to operate at 4~K~\cite{Homulle2017}.

\section{MGD Tuning and Stabilization Performance}
\label{sec:results}

\subsection{Minima Seeking}
\label{sec:scaling}

To evaluate the hardware framework on a localized stabilization task, we simulate a post-coarse-tune environment where the system must navigate to an absolute minimum. In this context, stabilization is defined as the algorithm's ability to navigate a high-dimensional, cross-coupled charge stability landscape and firmly park the system state within the target minimum, overcoming the surrounding noise floor.

Local stabilization is characterized for an idling parking point modeled by \RevTwo{the high-order polynomial cost function of \eqref{eq:idle}}{S19}. The parameters are chosen to emulate the region between steep transition peaks positioned just a few millivolts apart \RevTwo{(Fig.~\ref{fig:charge_stability_diagram}c)}{S20}. This landscape presents a challenge: steep outer walls paired with a nearly flat inner basin. This topology requires the hardware to survive exploding gradients during the initial descent, and then reliably navigate vanishing gradients obscured by ambient noise near the target minimum (Fig.~\ref{fig:heatmap}a).

To evaluate the efficiency of the MGD framework for highly coupled, multi-dimensional parameter spaces, we compiled a 100-gate version of the same two-sided update and measured the iterations to convergence on the FPGA for $N_\mathrm{Gates}$ from 2 to 100. To stress-test the controller within this polynomial basin, we also swept the cross-capacitance ratio $\kappa$ from 2 to 50 (Fig.~\hlref{fig:heatmap}b), with three trials per setting.

\Figure[t](topskip=0pt, botskip=0pt, midskip=0pt)[width=8.5cm]{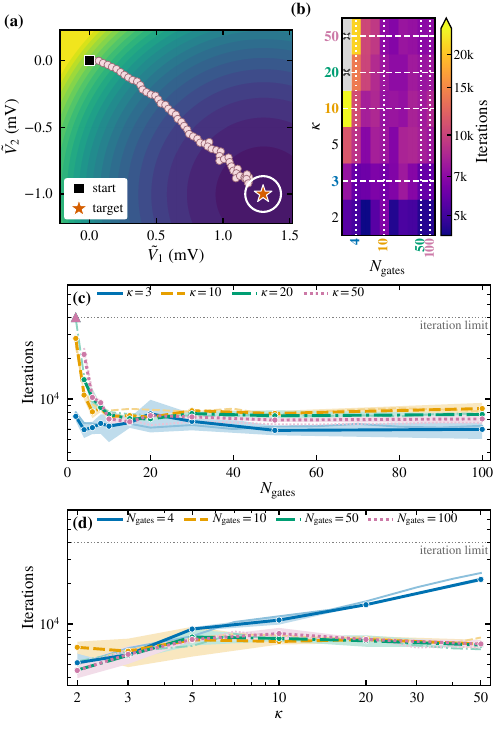}
{\textbf{MGD stabilization and performance with an increasing number of cross talking gates, measured on the FPGA.} (a) 2D trajectory (virtual gate 1 vs.\ 2) of the algorithm navigating the steep walls and crossing the flat basin to park within the target tolerance (circle). (b) Mean iterations to converge versus $N_\text{Gates}$ and $\kappa$ (3 trials per cell); grey crossed cells did not converge within $4\times10^{4}$ iterations. Dashed (dotted) lines mark the cuts in (c) [(d)]. (c) Horizontal cuts: the iterations show little increase with $N_\text{Gates}$ due to $\mathbf{A}$ being constrained to be rank-2. (d) Vertical cuts: iterations grow with $\kappa$ only for $N_\text{Gates}\leq 4$. Thick lines: FPGA, mean $\pm$ 1 s.d.; faint lines: software model.\label{fig:heatmap}}

The weak increase in run time with increasing numbers of physical gates (Fig.~\ref{fig:heatmap}c) follows directly from the facts that (i) because the number of measured dots is not scaling, there are still only two independent directions for all $N_\text{Gates}>2$ and (ii) the rows of the cross capacitance matrix $A$ are subject to exponential decay and thus adding \ReviewSpan{S21}{physical gates to the edge of the array has little to no impact}. \RevThree{This agrees with Ref.~\hlcite{oripov_scaling_2025}, which showed that training time in the MGD framework scales with the complexity of the task rather than with the number of parameters being optimized; here the task has two independent directions for any $N_\mathrm{Gates}$.}{S21} 

For $N_\mathrm{Gates} \geq 8$, every setting converged in $4.5$--$9.5\times10^{3}$ iterations, with no trend in $N_\mathrm{Gates}$ up to 100. The dependence on the condition number $\kappa$ (Fig.~\ref{fig:heatmap}d) is strong only when the dots couple to a few gates. For $N_\mathrm{Gates}=4$, the iterations rise by a factor of 4.1 between $\kappa=2$ and $\kappa=50$, and for $N_\mathrm{Gates}=2$ no trial converges within $4\times10^{4}$ iterations for $\kappa\geq20$. With only two gates, the narrow direction of the valley is reached only through the small difference between the two gates' couplings, so steps along it become small as the coupling grows. When the dots couple to eight or more gates, the dependence on $\kappa$ largely disappears: at $N_\mathrm{Gates}=10$ and 100, raising $\kappa$ from 5 to 20 changes the convergence time by a factor of 1.00, and from 20 to 50 by 0.93, compared with 2.0 and 1.58 for a $\kappa^{1/2}$ scaling. Each dot then responds to the combined perturbation of many neighboring gates, which appears to offset the narrowing of the valley. A software model of the update with the same cost, noise and convergence criteria reproduces the measurements (median ratio 0.98) and shows the same trends without noise, so they come from the landscape geometry rather than from the noise floor.
For the exponential cross-talk model used here,
$\kappa \approx \left[(1+r)/(1-r)\right]^2$, where $r = e^{-1/\gamma}$ is the
ratio of nearest-neighbor to direct gate coupling. Experimental devices
typically exhibit nearest-neighbor cross-talk of a few tens of percent
\cite{mills_computer-automated_2019}, corresponding to $\kappa \approx 2$--$9$ for $r$
between 0.2 and 0.5. The lower end of our sweep therefore represents
realistic devices, where even $N_\mathrm{Gates}=4$ converges within $1.1\times10^{4}$ iterations, while $\kappa = 50$ ($r \approx 0.75$) is a deliberate
stress test well beyond the coupling expected in practice.

\subsection{Noise Floor Failure Boundary}
\label{sec:noise}

To test the noise tolerance of our MGD implementation, the hardware was subjected to a noise intensity sweep for $N_\text{Gates}=10$. At each noise level, the white and $1/f$ components are scaled together, keeping their ratio at the Table~\hlref{tab:paramTable} values. The FPGA executed independent trials, with each trial granted a maximum of 40,000 iterations to converge before timing out. We consider the optimum reached when the distance from $\mathbf{V}_{\mathrm{opt}}$ in virtual-gate space, averaged over 250 iterations, falls below a tolerance of $134~\mu$V, 5\% of the 2.69~mV distance from the center of the cell to the transition peak. After a 20,000-iteration settling period, we record the fraction of time the controller stays within this tolerance, and count a trial as converged when it is acquired and held for at least 90\% of the remaining iterations.

The MGD algorithm is resilient to the slow components of the $1/f$ noise. Because the two measurements of an iteration are taken only 1~ms apart, the slowly varying octaves are nearly identical in both and largely cancel in the difference $C_+ - C_-$. The noise the controller acts on is dominated by the fastest octaves and the white noise: at the Table~\hlref{tab:paramTable} level for the measurement pairs the controller uses, its RMS is $3.3\times10^{-2}$ of the peak signal, compared with $6.0\times10^{-2}$ for a single measurement.

\Figure[!htbp](topskip=0pt, botskip=0pt, midskip=0pt)[width=8.5cm]{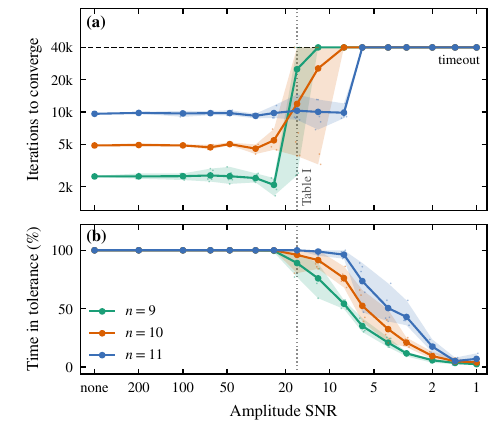}
{\textbf{Noise tolerance of the FPGA controller.} (a) Iterations to converge versus amplitude SNR for learning rates $n = 9$, 10 and 11 (lines: mean of five trials; points: individual trials). A trial counts as converged when it reaches the 134~$\mu$V tolerance and stays within it for at least 90\% of the scored iterations; otherwise it is plotted at the 40,000-iteration timeout. (b) Fraction of time within tolerance after a 20,000-iteration settling period. The dotted line marks the noise level of Table~\hlref{tab:paramTable} (SNR = 16.7). Each step down in $n$ halves the noise-free convergence time but moves the breakdown to a cleaner signal.\label{fig:noise}}

The noise sweep (Fig.~\ref{fig:noise}) reveals a trade-off between convergence speed and noise tolerance set entirely by the bit shift parameter $n$. Without noise, the controller converges in about 2,500, 4,900, and 9,600 iterations for $n = 9$, 10 and 11: each step down in $n$ doubles the step size and halves the convergence time. The larger steps, however, also amplify each noisy measurement, so the faster settings break down at cleaner signals. The mean convergence time reaches the timeout at an SNR of about 12 for $n = 9$, 8 for $n = 10$ and 6 for $n = 11$. At the noise level of Table~\hlref{tab:paramTable} (SNR = 16.7), all three settings acquire the operating point and hold it within tolerance 89\%, 96\% and 100\% of the time, respectively, although for $n = 9$ only two of five trials meet the 90\% criterion. Beyond the breakdown, the time in tolerance falls smoothly and reaches a few percent at SNR = 1, where the noise RMS equals the full transition peak signal. In practice, $n$ can therefore be chosen for speed in a quiet setup, or for robustness when the noise is larger than expected.

\subsection{Tracking a drifting operating point}
\label{sec:drift}

\Figure[t](topskip=0pt, botskip=0pt, midskip=0pt)[width=8.5cm]{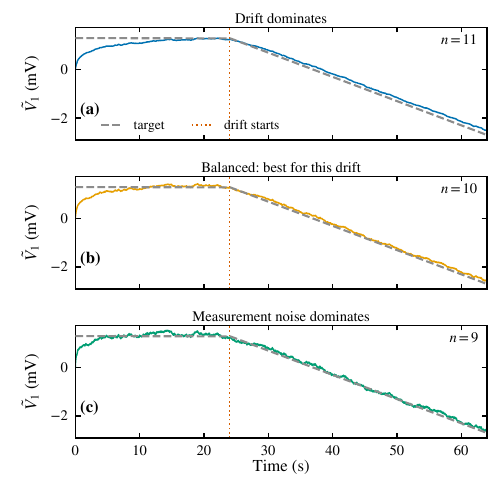}
{\textbf{Tracking a drifting operating point.} \ReviewSpan{S22}{(a)--(c) Virtual gate 1 versus time for $n = 11$, 10 and 9 while the idle point drifts at 0.1~mV/s per axis, at the noise level of Table~\hlref{tab:paramTable} (SNR = 16.7). The controller first acquires the idle point from 1.6~mV away; the target (dashed) starts drifting at the dotted line. The title of each panel names the effect that limits that setting. (a) With the smallest steps, $n = 11$ falls behind the drift, with a mean lag of 114~$\mu$V, and stays within 134~$\mu$V of the moving target 59\% of the time. (b) $n = 10$ balances the two effects, with a lag of 59~$\mu$V and 90\% of the time within tolerance. (c) With the largest steps, $n = 9$ follows the drift closely, with a lag of only 29~$\mu$V, but its jitter from measurement noise limits it to 83\% of the time within tolerance.}\label{fig:drift}}
To evaluate the algorithm's ability to track a moving operating point while rejecting noise, we subject the controller to two instabilities common in quantum dot arrays: a continuous drift of the operating point (this section) and discrete charge jumps (Section~\hlref{sec:jumps}).  Both tests include the noise of Table~\hlref{tab:paramTable}.

 In the drift test, the controller first acquires the idling point of \eqref{eq:idle} from 1.6~mV away; after 12,000 iterations (24~s, as each iteration takes two 1~ms measurements), $\mathbf{V}_\text{opt}$ begins to move at a constant rate $R$ along both virtual-gate axes. We quantify tracking by the fraction of iterations the controller stays within a threshold $V_{\mathrm{th}} = 134~\mu$V of the moving target, the same tolerance used in Section~\ref{sec:noise},
\begin{equation}
    \text{Eff}(V_\text{th}) = \frac{1}{N_\text{total}}\sum_{k=1}^{N_\text{total}}\left[ \left|\left|\mathbf{V}_k-\mathbf{V}_\text{opt}(t_k)\right|\right| \leq V_\text{th}\right]\times100\%,
    \label{eq:eff}
\end{equation}
\noindent where $N_\text{total}$ is the number of iterations scored, counted from 2,000 iterations (4~s) after the drift begins, and the distance is measured in virtual-gate space. The rates tested, 0.03 to 0.3~mV/s, are far faster than the drifts observed in real devices ~\cite{puddy_multiplexed_2015, rao_towards_2025}.

The learning rate sets a trade-off that mirrors the noise sweep of Section~\hlref{sec:noise} (Fig.~\hlref{fig:drift}). At $R = 0.03$~mV/s, $n = 10$ and $n = 11$ stay within tolerance for 100\% of the drift and $n = 9$ for 88\%, limited by its larger jitter. At $R = 0.1$~mV/s, the slower settings fall behind: the mean lag grows from 29~$\mu$V for $n = 9$ to 59 and 114~$\mu$V for $n = 10$ and 11, which stay within tolerance 83\%, 90\% and 59\% of the time, respectively. At $R = 0.3$~mV/s, only $n = 9$ keeps up, staying within tolerance 64\% of the time, while $n = 10$ and 11 lag by more than the tolerance. A smaller $n$ therefore follows faster drifts at the cost of noise tolerance, and \ReviewSpan{S23}{the best setting depends on whether the drift or the measurement noise dominates}. \RevThree{The iteration rate is set by the readout: each iteration here takes two 1~ms measurements, while the FPGA logic adds only tens of nanoseconds. A faster readout lets the controller follow proportionally faster drifts at the same $n$, and a slower one lowers the drift rate it can track, so the measurement speed rather than the controller sets this limit. A shorter measurement, however, generally carries more white noise, which moves the optimum toward larger $n$.}{S23}

\subsection{Recovery from charge jumps}
\label{sec:jumps}

\Figure[t](topskip=0pt, botskip=0pt, midskip=0pt)[width=8.5cm]{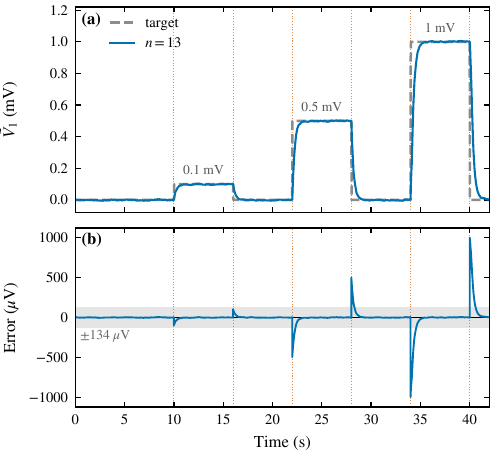}
{\textbf{Recovery from charge jumps at the pulsing point.} (a) Virtual gate 1 (solid) following the target (dashed), which jumps by 0.1, 0.5 and 1~mV and returns (dotted lines), for $n = 13$ at the noise level of Table~\hlref{tab:paramTable}. \ReviewSpan{S24}{(b) Deviation from the target}; the shaded band is the 134~$\mu$V tolerance.\label{fig:jumps}}

To test recovery from discrete charge-switching events, we hold the pulsing point using the cost of \eqref{eq:pulse} and shift $\mathbf{V}_\text{opt}$ by 0.1, 0.5 and 1~mV along both virtual-gate axes, returning it after 3,000 iterations (6~s) each time. With $n = 13$, the controller returns to within 134~$\mu$V of the target after about 20~ms, 0.4~s and 0.7~s (10, 200 and 330 iterations) for the three jumps, and stays within tolerance 95\% of the time overall (Fig.~\ref{fig:jumps}). For jumps larger than $\epsilon$, the measured cost difference is at its maximum and the gates first move at a constant rate; within $\epsilon$ of the target, the two measurement points straddle the minimum, the measured slope becomes proportional to the remaining error, and the error decays smoothly to zero. The controller therefore settles without oscillating about the minimum, and no deadband is required. Between jumps, the operating point is held with an RMS deviation of only 3~$\mu$V.


\section{Summary and Outlook}
\label{sec:summary}

In this work we have presented a robust, automatic, and device-agnostic method for stabilizing quantum dot qubits at optimal operating points in spite of noise, drift, and jumps in the potential landscape. The algorithm is found to converge quickly to an optimal idling point in spite of a high dimensional parameter space and a noisy environment. Adding weakly coupled gates around the double dot has little effect on the convergence time, since the cost depends only  on the two virtual-gate directions of the dots. Additionally, the algorithm proves able to rest on the steep-gradient flank where there is greater noise and current through the dot, in spite of drifts and discrete jumps in the charge stability diagram.

Moreover, this automated control method bypasses any need for repeated global scans of the charge stability diagram, opening the door for future work to scale this solution in the face of an ever-expanding voltage parameter space. Furthermore, the FPGA implementation of this tuning algorithm allows for future cryogenic implementation on a 4 K stage. This paves the way for future work, where large quantum dot arrays can be automatically stabilized at optimal control points completely \textit{in-situ}.

\section*{Acknowledgment}
Claude LLM was used in preparing the noise sweep routine of Sec. \ref{sec:noise}, as well as for plotting, grammar, and editing. All LLM output has been independently verified for correctness by the authors.

\bibliographystyle{IEEEtran}
\bibliography{references}

@article{zwolak_colloquium_2023,
    title = {Colloquium: {Advances} in automation of quantum dot devices control},
    volume = {95},
    shorttitle = {Colloquium},
    url = {https://link.aps.org/doi/10.1103/RevModPhys.95.011006},
    doi = {10.1103/RevModPhys.95.011006},
    number = {1},
    urldate = {2026-06-15},
    journal = {Reviews of Modern Physics},
    publisher = {American Physical Society},
    author = {Zwolak, Justyna P. and Taylor, Jacob M.},
    month = feb,
    year = {2023},
    pages = {011006},
}

@article{schuff_fully_2026,
    title = {Fully autonomous tuning of a spin qubit},
    volume = {9},
    copyright = {2026 The Author(s)},
    issn = {2520-1131},
    url = {https://www.nature.com/articles/s41928-025-01562-4},
    doi = {10.1038/s41928-025-01562-4},
    number = {3},
    urldate = {2026-06-15},
    journal = {Nature Electronics},
    publisher = {Nature Publishing Group},
    author = {Schuff, Jonas and Carballido, Miguel J. and Kotzagiannidis, Madeleine and Calvo, Juan Carlos and Caselli, Marco and Rawling, Jacob and Craig, David L. and van Straaten, Barnaby and Severin, Brandon and Fedele, Federico and Svab, Simon and Chevalier Kwon, Pierre and Eggli, Rafael S. and Patlatiuk, Taras and Korda, Nathan and Zumbühl, Dominik M. and Ares, Natalia},
    month = mar,
    year = {2026},
    pages = {304--313},
}

@article{botzem_tuning_2018,
    title = {Tuning {Methods} for {Semiconductor} {Spin} {Qubits}},
    volume = {10},
    url = {https://link.aps.org/doi/10.1103/PhysRevApplied.10.054026},
    doi = {10.1103/PhysRevApplied.10.054026},
    number = {5},
    urldate = {2026-06-15},
    journal = {Physical Review Applied},
    publisher = {American Physical Society},
    author = {Botzem, Tim and Shulman, Michael D. and Foletti, Sandra and Harvey, Shannon P. and Dial, Oliver E. and Bethke, Patrick and Cerfontaine, Pascal and McNeil, Robert P. G. and Mahalu, Diana and Umansky, Vladimir and Ludwig, Arne and Wieck, Andreas and Schuh, Dieter and Bougeard, Dominique and Yacoby, Amir and Bluhm, Hendrik},
    month = nov,
    year = {2018},
    pages = {054026},
}

@misc{rao_towards_2025,
    title = {Towards autonomous time-calibration of large quantum-dot devices: {Detection}, real-time feedback, and noise spectroscopy},
    shorttitle = {Towards autonomous time-calibration of large quantum-dot devices},
    url = {http://arxiv.org/abs/2512.24894},
    doi = {10.48550/arXiv.2512.24894},
    urldate = {2026-06-15},
    publisher = {arXiv},
    author = {Rao, Anantha S. and Straaten, Barnaby van and John, Valentin and Yu, Cécile X. and Oosterhout, Stefan D. and Stehouwer, Lucas and Scappucci, Giordano and Stewart, M. D. and Veldhorst, Menno and Borsoi, Francesco and Zwolak, Justyna P.},
    month = dec,
    year = {2025},
    note = {arXiv:2512.24894 [cond-mat.mes-hall]},
}

@article{taylor_spin_2026,
  title={Spin qubit tuning automation for computer scientists},
  author={Taylor, Brady and Misra, Shashank and Ivie, Jeffrey A and Allemang, Christopher Richard},
  journal={Quantum Science and Technology},
  year={2026}
}

@article{dillavou_machine_2024,
    title = {Machine learning without a processor: {Emergent} learning in a nonlinear analog network},
    volume = {121},
    shorttitle = {Machine learning without a processor},
    url = {https://www.pnas.org/doi/10.1073/pnas.2319718121},
    doi = {10.1073/pnas.2319718121},
    number = {28},
    urldate = {2026-06-15},
    journal = {Proceedings of the National Academy of Sciences},
    publisher = {Proceedings of the National Academy of Sciences},
    author = {Dillavou, Sam and Beyer, Benjamin D. and Stern, Menachem and Liu, Andrea J. and Miskin, Marc Z. and Durian, Douglas J.},
    month = jul,
    year = {2024},
    pages = {e2319718121},
}

@article{mccaughan_multiplexed_2023,
    title = {Multiplexed gradient descent: {Fast} online training of modern datasets on hardware neural networks without backpropagation},
    volume = {1},
    issn = {2770-9019},
    shorttitle = {Multiplexed gradient descent},
    url = {https://doi.org/10.1063/5.0157645},
    doi = {10.1063/5.0157645},
    number = {2},
    urldate = {2026-06-15},
    journal = {APL Machine Learning},
    author = {McCaughan, Adam N. and Oripov, Bakhrom G. and Ganesh, Natesh and Nam, Sae Woo and Dienstfrey, Andrew and Buckley, Sonia M.},
    month = jun,
    year = {2023},
    pages = {026118},
}

@article{paladino_mathbsf1mathbsfitf_2014,
    title = {${\mathbf{1}}/{\mathit{f}}$ noise: {Implications} for solid-state quantum information},
    volume = {86},
    shorttitle = {\${\textbackslash}mathbsf\{1\}/{\textbackslash}mathbsfit\{f\}\$ noise},
    url = {https://link.aps.org/doi/10.1103/RevModPhys.86.361},
    doi = {10.1103/RevModPhys.86.361},
    number = {2},
    urldate = {2026-06-15},
    journal = {Reviews of Modern Physics},
    publisher = {American Physical Society},
    author = {Paladino, E. and Galperin, Y. M. and Falci, G. and Altshuler, B. L.},
    month = apr,
    year = {2014},
    pages = {361--418},
}

@article{dial_charge_2013,
    title = {Charge {Noise} {Spectroscopy} {Using} {Coherent} {Exchange} {Oscillations} in a {Singlet}-{Triplet} {Qubit}},
    volume = {110},
    url = {https://link.aps.org/doi/10.1103/PhysRevLett.110.146804},
    doi = {10.1103/PhysRevLett.110.146804},
    number = {14},
    urldate = {2026-06-15},
    journal = {Physical Review Letters},
    publisher = {American Physical Society},
    author = {Dial, O. E. and Shulman, M. D. and Harvey, S. P. and Bluhm, H. and Umansky, V. and Yacoby, A.},
    month = apr,
    year = {2013},
    pages = {146804},
}

@article{kalantre_machine_2019,
    title = {Machine learning techniques for state recognition and auto-tuning in quantum dots},
    volume = {5},
    copyright = {2019 The Author(s)},
    issn = {2056-6387},
    url = {https://www.nature.com/articles/s41534-018-0118-7},
    doi = {10.1038/s41534-018-0118-7},
    number = {1},
    urldate = {2026-06-15},
    journal = {npj Quantum Information},
    publisher = {Nature Publishing Group},
    author = {Kalantre, Sandesh S. and Zwolak, Justyna P. and Ragole, Stephen and Wu, Xingyao and Zimmerman, Neil M. and Stewart, M. D. and Taylor, Jacob M.},
    month = jan,
    year = {2019},
    pages = {6},
}

@article{teske_machine_2019,
    title = {A machine learning approach for automated fine-tuning of semiconductor spin qubits},
    volume = {114},
    issn = {0003-6951},
    url = {https://doi.org/10.1063/1.5088412},
    doi = {10.1063/1.5088412},
    number = {13},
    urldate = {2026-06-15},
    journal = {Applied Physics Letters},
    author = {Teske, Julian D. and Humpohl, Simon Sebastian and Otten, René and Bethke, Patrick and Cerfontaine, Pascal and Dedden, Jonas and Ludwig, Arne and Wieck, Andreas D. and Bluhm, Hendrik},
    month = apr,
    year = {2019},
    pages = {133102},
}

@article{loss_quantum_1998,
    title = {Quantum computation with quantum dots},
    volume = {57},
    url = {https://link.aps.org/doi/10.1103/PhysRevA.57.120},
    doi = {10.1103/PhysRevA.57.120},
    number = {1},
    urldate = {2026-06-23},
    journal = {Physical Review A},
    publisher = {American Physical Society},
    author = {Loss, Daniel and DiVincenzo, David P.},
    month = jan,
    year = {1998},
    pages = {120--126},
}

@article{yoneda_quantum-dot_2018,
    title = {A quantum-dot spin qubit with coherence limited by charge noise and fidelity higher than 99.9\%},
    volume = {13},
    copyright = {2017 The Author(s)},
    issn = {1748-3395},
    url = {https://www.nature.com/articles/s41565-017-0014-x},
    doi = {10.1038/s41565-017-0014-x},
    
    number = {2},
    urldate = {2026-06-23},
    journal = {Nature Nanotechnology},
    publisher = {Nature Publishing Group},
    author = {Yoneda, Jun and Takeda, Kenta and Otsuka, Tomohiro and Nakajima, Takashi and Delbecq, Matthieu R. and Allison, Giles and Honda, Takumu and Kodera, Tetsuo and Oda, Shunri and Hoshi, Yusuke and Usami, Noritaka and Itoh, Kohei M. and Tarucha, Seigo},
    month = feb,
    year = {2018},
    pages = {102--106},
}

@article{voss_1f_1978,
    title = {``1/f noise'' in music: {Music} from 1/f noise},
    volume = {63},
    issn = {0001-4966},
    shorttitle = {``1/f noise'' in music},
    url = {https://doi.org/10.1121/1.381721},
    doi = {10.1121/1.381721},
    number = {1},
    urldate = {2026-06-30},
    journal = {The Journal of the Acoustical Society of America},
    author = {Voss, Richard F. and Clarke, John},
    month = jan,
    year = {1978},
    pages = {258--263},
}

@article{connors_charge-noise_2022,
    title = {Charge-noise spectroscopy of {Si}/{SiGe} quantum dots via dynamically-decoupled exchange oscillations},
    volume = {13},
    copyright = {2022 The Author(s)},
    issn = {2041-1723},
    url = {https://www.nature.com/articles/s41467-022-28519-x},
    doi = {10.1038/s41467-022-28519-x},
    
    number = {1},
    urldate = {2026-06-29},
    journal = {Nature Communications},
    publisher = {Nature Publishing Group},
    author = {Connors, Elliot J. and Nelson, J. and Edge, Lisa F. and Nichol, John M.},
    month = feb,
    year = {2022},
    pages = {940},
}

@article{paquelet_wuetz_reducing_2023,
    title = {Reducing charge noise in quantum dots by using thin silicon quantum wells},
    volume = {14},
    copyright = {2023 The Author(s)},
    issn = {2041-1723},
    url = {https://www.nature.com/articles/s41467-023-36951-w},
    doi = {10.1038/s41467-023-36951-w},
    
    number = {1},
    urldate = {2026-06-29},
    journal = {Nature Communications},
    publisher = {Nature Publishing Group},
    author = {Paquelet Wuetz, Brian and Degli Esposti, Davide and Zwerver, Anne-Marije J. and Amitonov, Sergey V. and Botifoll, Marc and Arbiol, Jordi and Sammak, Amir and Vandersypen, Lieven M. K. and Russ, Maximilian and Scappucci, Giordano},
    month = mar,
    year = {2023},
    pages = {1385},
}

@article{Homulle2017,
  author  = {Homulle, Harald and Visser, Stefan and Patra, Bishnu and Ferrari, Giorgio and Prati, Enrico and Sebastiano, Fabio and Charbon, Edoardo},
  title   = {A reconfigurable cryogenic platform for the classical control of quantum processors},
  journal = {Review of Scientific Instruments},
  volume  = {88},
  number  = {4},
  pages   = {045103},
  year    = {2017},
  doi     = {10.1063/1.4979611}
}

@article{elsayed_low_2024,
    title = {Low charge noise quantum dots with industrial {CMOS} manufacturing},
    volume = {10},
    copyright = {2024 The Author(s)},
    issn = {2056-6387},
    url = {https://www.nature.com/articles/s41534-024-00864-3},
    doi = {10.1038/s41534-024-00864-3},
    
    number = {1},
    urldate = {2026-06-30},
    journal = {npj Quantum Information},
    publisher = {Nature Publishing Group},
    author = {Elsayed, A. and Shehata, M. M. K. and Godfrin, C. and Kubicek, S. and Massar, S. and Canvel, Y. and Jussot, J. and Simion, G. and Mongillo, M. and Wan, D. and Govoreanu, B. and Radu, I. P. and Li, R. and Van Dorpe, P. and De Greve, K.},
    month = jul,
    year = {2024},
    pages = {70},
}

@article{koch_industrial_2025,
    title = {Industrial 300 mm wafer processed spin qubits in natural silicon/silicon-germanium},
    volume = {11},
    copyright = {2025 The Author(s)},
    issn = {2056-6387},
    url = {https://www.nature.com/articles/s41534-025-01016-x},
    doi = {10.1038/s41534-025-01016-x},
    
    number = {1},
    urldate = {2026-07-06},
    journal = {npj Quantum Information},
    publisher = {Nature Publishing Group},
    author = {Koch, Thomas and Godfrin, Clement and Adam, Viktor and Ferrero, Julian and Schroller, Daniel and Glaeser, Noah and Kubicek, Stefan and Li, Ruoyu and Loo, Roger and Massar, Shana and Simion, George and Wan, Danny and De Greve, Kristiaan and Wernsdorfer, Wolfgang},
    month = apr,
    year = {2025},
    pages = {59},
}

@article{paquelet_wuetz_atomic_2022,
    title = {Atomic fluctuations lifting the energy degeneracy in {Si}/{SiGe} quantum dots},
    volume = {13},
    copyright = {2022 The Author(s)},
    issn = {2041-1723},
    url = {https://www.nature.com/articles/s41467-022-35458-0},
    doi = {10.1038/s41467-022-35458-0},
    number = {1},
    urldate = {2026-07-06},
    journal = {Nature Communications},
    publisher = {Nature Publishing Group},
    author = {Paquelet Wuetz, Brian and Losert, Merritt P. and Koelling, Sebastian and Stehouwer, Lucas E. A. and Zwerver, Anne-Marije J. and Philips, Stephan G. J. and Mądzik, Mateusz T. and Xue, Xiao and Zheng, Guoji and Lodari, Mario and Amitonov, Sergey V. and Samkharadze, Nodar and Sammak, Amir and Vandersypen, Lieven M. K. and Rahman, Rajib and Coppersmith, Susan N. and Moutanabbir, Oussama and Friesen, Mark and Scappucci, Giordano},
    month = dec,
    year = {2022},
    pages = {7730},
}

@article{puddy_multiplexed_2015,
    title = {Multiplexed charge-locking device for large arrays of quantum devices},
    volume = {107},
    issn = {0003-6951},
    url = {https://doi.org/10.1063/1.4932012},
    doi = {10.1063/1.4932012},
    number = {14},
    urldate = {2026-07-06},
    journal = {Applied Physics Letters},
    author = {Puddy, R. K. and Smith, L. W and Al-Taie, H. and Chong, C. H. and Farrer, I. and Griffiths, J. P. and Ritchie, D. A. and Kelly, M. J. and Pepper, M. and Smith, C. G.},
    month = oct,
    year = {2015},
    pages = {143501},
}

@article{rudolph_long-term_2019,
    title = {Long-term drift of {Si}-{MOS} quantum dots with intentional donor implants},
    volume = {9},
    copyright = {2019 The Author(s)},
    issn = {2045-2322},
    url = {https://www.nature.com/articles/s41598-019-43995-w},
    doi = {10.1038/s41598-019-43995-w},
    number = {1},
    urldate = {2026-07-06},
    journal = {Scientific Reports},
    publisher = {Nature Publishing Group},
    author = {Rudolph, M. and Sarabi, B. and Murray, R. and Carroll, M. S. and Zimmerman, Neil M.},
    month = may,
    year = {2019},
    pages = {7656},
}

@article{volk_loading_2019,
    title = {Loading a quantum-dot based “{Qubyte}” register},
    volume = {5},
    copyright = {2019 The Author(s)},
    issn = {2056-6387},
    url = {https://www.nature.com/articles/s41534-019-0146-y},
    doi = {10.1038/s41534-019-0146-y},
    number = {1},
    urldate = {2026-07-08},
    journal = {npj Quantum Information},
    publisher = {Nature Publishing Group},
    author = {Volk, C. and Zwerver, A. M. J. and Mukhopadhyay, U. and Eendebak, P. T. and van Diepen, C. J. and Dehollain, J. P. and Hensgens, T. and Fujita, T. and Reichl, C. and Wegscheider, W. and Vandersypen, L. M. K.},
    month = apr,
    year = {2019},
    pages = {29},
}

@article{moon_machine_2020,
    title = {Machine learning enables completely automatic tuning of a quantum device faster than human experts},
    volume = {11},
    copyright = {2020 The Author(s)},
    issn = {2041-1723},
    url = {https://www.nature.com/articles/s41467-020-17835-9},
    doi = {10.1038/s41467-020-17835-9},
    language = {en},
    number = {1},
    urldate = {2026-08-19},
    journal = {Nature Communications},
    publisher = {Nature Publishing Group},
    author = {Moon, H. and Lennon, D. T. and Kirkpatrick, J. and van Esbroeck, N. M. and Camenzind, L. C. and Yu, Liuqi and Vigneau, F. and Zumbühl, D. M. and Briggs, G. a. D. and Osborne, M. A. and Sejdinovic, D. and Laird, E. A. and Ares, N.},
    month = aug,
    year = {2020},
    pages = {4161},
}

@article{baart_computer-automated_2016,
    title = {Computer-automated tuning of semiconductor double quantum dots into the single-electron regime},
    volume = {108},
    issn = {0003-6951},
    url = {https://doi.org/10.1063/1.4952624},
    doi = {10.1063/1.4952624},
    number = {21},
    urldate = {2026-09-23},
    journal = {Applied Physics Letters},
    author = {Baart, T. A. and Eendebak, P. T. and Reichl, C. and Wegscheider, W. and Vandersypen, L. M. K.},
    month = may,
    year = {2016},
    pages = {213104},
}

@article{durrer_automated_2020,
    title = {Automated {Tuning} of {Double} {Quantum} {Dots} into {Specific} {Charge} {States} {Using} {Neural} {Networks}},
    volume = {13},
    url = {https://link.aps.org/doi/10.1103/PhysRevApplied.13.054019},
    doi = {10.1103/PhysRevApplied.13.054019},
    number = {5},
    urldate = {2026-08-19},
    journal = {Physical Review Applied},
    publisher = {American Physical Society},
    author = {Durrer, R. and Kratochwil, B. and Koski, J.V. and Landig, A.J. and Reichl, C. and Wegscheider, W. and Ihn, T. and Greplova, E.},
    month = may,
    year = {2020},
    pages = {054019},
}

@article{mills_computer-automated_2019,
    title = {Computer-automated tuning procedures for semiconductor quantum dot arrays},
    volume = {115},
    issn = {0003-6951},
    url = {https://doi.org/10.1063/1.5121444},
    doi = {10.1063/1.5121444},
    number = {11},
    urldate = {2026-09-23},
    journal = {Applied Physics Letters},
    author = {Mills, A. R. and Feldman, M. M. and Monical, C. and Lewis, P. J. and Larson, K. W. and Mounce, A. M. and Petta, J. R.},
    month = sep,
    year = {2019},
    pages = {113501},
}

@article{van_diepen_automated_2018,
    title = {Automated tuning of inter-dot tunnel coupling in double quantum dots},
    volume = {113},
    issn = {0003-6951},
    url = {https://doi.org/10.1063/1.5031034},
    doi = {10.1063/1.5031034},
    number = {3},
    urldate = {2026-09-23},
    journal = {Applied Physics Letters},
    author = {van Diepen, C. J. and Eendebak, P. T. and Buijtendorp, B. T. and Mukhopadhyay, U. and Fujita, T. and Reichl, C. and Wegscheider, W. and Vandersypen, L. M. K.},
    month = jul,
    year = {2018},
    pages = {033101},
}

@article{nakajima_real-time_2021,
    title = {Real-{Time} {Feedback} {Control} of {Charge} {Sensing} for {Quantum} {Dot} {Qubits}},
    volume = {15},
    url = {https://link.aps.org/doi/10.1103/PhysRevApplied.15.L031003},
    doi = {10.1103/PhysRevApplied.15.L031003},
    number = {3},
    urldate = {2026-09-23},
    journal = {Physical Review Applied},
    publisher = {American Physical Society},
    author = {Nakajima, Takashi and Kojima, Yohei and Uehara, Yoshihiro and Noiri, Akito and Takeda, Kenta and Kobayashi, Takashi and Tarucha, Seigo},
    month = mar,
    year = {2021},
    pages = {L031003},
}

@article{oripov_scaling_2025,
    title = {Scaling of hardware-compatible perturbative training algorithms},
    volume = {3},
    issn = {2770-9019},
    url = {https://doi.org/10.1063/5.0258271},
    doi = {10.1063/5.0258271},
    number = {2},
    urldate = {2026-09-23},
    journal = {APL Machine Learning},
    author = {Oripov, B. G. and Dienstfrey, A. and McCaughan, A. N. and Buckley, S. M.},
    month = apr,
    year = {2025},
    pages = {026107},
}

@inproceedings{spall_stochastic_1987,
    title = {A {Stochastic} {Approximation} {Technique} for {Generating} {Maximum} {Likelihood} {Parameter} {Estimates}},
    url = {https://ieeexplore.ieee.org/abstract/document/4789489},
    doi = {10.23919/ACC.1987.4789489},
    urldate = {2026-09-23},
    booktitle = {1987 {American} {Control} {Conference}},
    author = {Spall, James C.},
    month = jun,
    year = {1987},
    pages = {1161--1167},
}

@incollection{van_houten_coulomb-blockade_1992,
    address = {Boston, MA},
    title = {Coulomb-{Blockade} {Oscillations} in {Semiconductor} {Nanostructures}},
    isbn = {978-1-4757-2166-9},
    url = {https://doi.org/10.1007/978-1-4757-2166-9_5},
    doi = {10.1007/978-1-4757-2166-9_5},
    booktitle = {Single {Charge} {Tunneling}: {Coulomb} {Blockade} {Phenomena} {In} {Nanostructures}},
    publisher = {Springer US},
    author = {Van Houten, H. and Beenakker, C. W. J. and Staring, A. A. M.},
    editor = {Grabert, Hermann and Devoret, Michel H.},
    year = {1992},
    pages = {167--216},
}

@article{krstic_stability_2000,
  title = {Stability of extremum seeking feedback for general nonlinear dynamic systems},
  volume = {36},
  issn = {0005-1098},
  url = {https://www.sciencedirect.com/science/article/pii/S0005109899001831},
  doi = {10.1016/S0005-1098(99)00183-1},
  number = {4},
  urldate = {2026-10-26},
  journal = {Automatica},
  author = {Krstić, Miroslav and Wang, Hsin-Hsiung},
  month = apr,
  year = {2000},
  pages = {595--601},
}

@book{liu_stochastic_2012,
  address = {London},
  series = {Communications and {Control} {Engineering}},
  title = {Stochastic {Averaging} and {Stochastic} {Extremum} {Seeking}},
  copyright = {https://www.springernature.com/gp/researchers/text-and-data-mining},
  isbn = {978-1-4471-4086-3 978-1-4471-4087-0},
  url = {https://link.springer.com/10.1007/978-1-4471-4087-0},
  doi = {10.1007/978-1-4471-4087-0},
  language = {en},
  urldate = {2026-10-26},
  publisher = {Springer},
  author = {Liu, Shu-Jun and Krstic, Miroslav},
  year = {2012},
}

@article{scheinker_100_2024,
  title = {100 years of extremum seeking: {A} survey},
  volume = {161},
  issn = {0005-1098},
  shorttitle = {100 years of extremum seeking},
  url = {https://doi.org/10.1016/j.automatica.2023.111481},
  doi = {10.1016/j.automatica.2023.111481},
  number = {C},
  urldate = {2026-10-26},
  journal = {Automatica},
  author = {Scheinker, Alexander},
  month = mar,
  year = {2024},
}

@article{hader_simulation_2024,
  title = {Simulation of {Charge} {Stability} {Diagrams} for {Automated} {Tuning} {Solutions} ({SimCATS})},
  volume = {5},
  issn = {2689-1808},
  url = {https://ieeexplore.ieee.org/abstract/document/10640289},
  doi = {10.1109/TQE.2024.3445967},
  urldate = {2026-11-06},
  journal = {IEEE Transactions on Quantum Engineering},
  author = {Hader, Fabian and Fleitmann, Sarah and Vogelbruch, Jan and Geck, Lotte and Waasen, Stefan van},
  year = {2024},
  pages = {1--14},
}

% ...

\EOD

\end{document}